\documentclass[aps,twocolumn,tightenlines,showpacs]{revtex4}
\usepackage[dvipdf]{graphicx}
\usepackage{color}
\usepackage{amssymb,amsmath,amsbsy}

\begin{document}

\title{Classical analogue of the continuous transition between \\ the Weisskopf-Wigner exponential decay and the Rabi oscillation}

\author{Gilles Dolfo}
\email{gilles.dolfo@irsamc.ups-tlse.fr}

\author{Jacques Vigu\'e}

\affiliation{ Laboratoire Collisions Agr\'egats R\'eactivit\'e-IRSAMC
\\Universit\'e de Toulouse-UPS and CNRS UMR 5589
\\ 118, Route de Narbonne, 31062 Toulouse Cedex, France}

\date{\today}

\begin{abstract}

When a discrete state is coupled to a continuum, the dynamics can be described either by the Weisskopf-Wigner exponential decay or by the Rabi oscillation, depending on the relative magnitudes of the continuum width and of the Rabi frequency. A continuous transition between these two regimes exists, as demonstrated in 1977 by C. Cohen-Tannoudji and P. Avan. Here, we describe a fully analogous transition in classical mechanics, by studying the dynamics of two coupled mechanical oscillators in the presence of damping. By varying the relative magnitudes of the damping and coupling terms, we observe a continuous transition between a regime analogous to the Rabi oscillation and a regime analogous to the Weisskopf-Wigner exponential decay.

\end{abstract}

\maketitle

\section{Introduction}
\label{se1}

It is usually very fruitful to explore the connection between classical and quantum mechanics but this connection is almost always done by going from classical mechanics toward quantum mechanics. Here, we go the other way round and exhibit the classical analogue of the transition between the Weisskopf-Wigner exponential decay \cite{WW1930a,WW1930b} and the Rabi oscillation \cite{RabiPR37}: these two quantum mechanics regimes are very well known but it is only in 1977 that the existence of a continuous transition between them was exhibited by C. Cohen-Tannoudji and P. Avan \cite{Cohentannoudji77}. 

This transition is described in detail in the book ``Atom-photon interactions'', by C. Cohen-Tannoudji, J. Dupont-Roc and G. Grynberg \cite{Cohentannoudji88}. At first sight, these two regimes seem to be very different because the Weisskopf-Wigner exponential decay appears when a discrete state is coupled to a continuum while the Rabi oscillation occurs when two discrete states are resonantly coupled. However, in the presence of the radiation continuum, any atomic system has only one discrete state, its ground state (i.e. the atomic ground state and the radiation vacuum state), and all the excited atomic states are narrow continua with a width equal to their radiative natural width. As a consequence, the Rabi oscillation between the atomic ground and an excited state, treated as a discrete state, is an approximation of the real situation because the excited state is in fact a narrow continuum: this approximation is excellent for times shorter than the excited state lifetime and Rabi oscillation is observed if the Rabi period is shorter than the excited state lifetime. The Weisskopf-Wigner exponential decay is an excellent approximation in the opposite limit, when the Rabi period is larger than the excited state lifetime.

In classical mechanics textbooks, the dynamics of two coupled mechanical oscillators is usually discussed without damping terms. In this case, the frequencies of the coupled oscillators present an avoided crossing, which is fully analogous to the avoided crossings of the eigenvalues of a Hamiltonian \cite{Landau65} and the dynamics, with a periodic exchange of energy between the two oscillators, is fully analogous to the Rabi oscillation. However, in the absence of damping terms, it is not possible to observe the classical analogue of the Weisskopf-Wigner exponential decay. Some textbooks on mechanical vibrations take into account damping in their treatment of coupled oscillators: this is the case of the book ``Mechanical vibrations'', by J.P. Den Hartog \cite{DenHartog56} which studies the damping of an oscillator by coupling to another oscillator. This device, patented by H. Frahm \cite{Frahm11} in 1911, is now known as a ``tuned mass damper'' and it has many applications. However, this book does not consider the general case discussed here.   

In the present paper, we study theoretically two coupled oscillators with damping and we observe two limiting cases:

i) if the coupling effect is dominant in a sense explained in section \ref{se4}, the mixing of the two oscillators is not substantially modified by the presence of damping. In particular, this mixing induces an averaging of the damping rates which are equal in the case of exact resonance. In this case, the dynamics remains an analogue of the Rabi oscillation;

ii) if the damping effect is dominant, the mixing of the two oscillators is strongly modified by the presence of damping. In particular, if one of the two oscillators has a negligible damping rate, the situation is completely analogous to the coupling of a discrete state to a continuum: the damping rate of this oscillator, which is only due to its mixing with the other oscillator, decreases if the damping rate of the other oscillator increases. A similar result is observed in the Weisskopf-Wigner model, with the decay rate decreasing when the continuum width increases. The resonant frequency of this oscillator also presents variations, which are fully analogous to the frequency shift of a discrete level due to its coupling to a continuum.

The content of the present paper is organized as follows: section \ref{se2} recalls the coupling of two mechanical oscillators without damping; the same problem with damping is discussed in section \ref{se3}; section \ref{se4} presents some concluding remarks. Two appendices recall Newton's equations of two coupled mechanical oscillators (section \ref{se5}) and general properties concerning the damping of mechanical oscillators (section \ref{se6}).

\section{Coupled mechanical oscillators without damping}
\label{se2}

\subsection{Calculation of the oscillation frequencies}

\begin{figure}[h]
\begin{center}
\includegraphics[width= 8 cm]{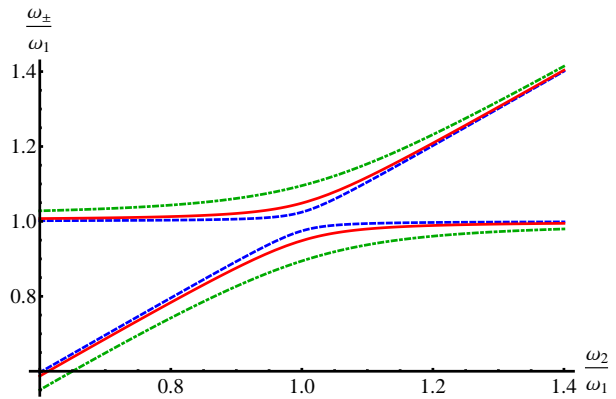}
\caption{The values of $\omega_{\pm}/\omega_1$ plotted as a function of the ratio $\omega_2/\omega_1$ exhibit an avoided crossing with a width proportional to the coupling constant $\kappa$.  The curves correspond to $\kappa= 0.05 $ dashed (blue) curves $\kappa=0.1$ full (red) curves and  $\kappa=0.2$ dot-dashed (green) curves.}
\label{fig1}
\end{center}
\end{figure}

Newton's equations of two coupled pendulums or a double pendulum (see Appendix A) take the same form 
\begin{eqnarray}\label{s0}
\frac{d^{2}x_1}{dt^{2}} &=& -\omega_1^{2} x_1 +\omega_{12}^{2} x_2, \nonumber \\
\frac{d^{2}x_2}{dt^{2}} &=&  +\omega_{21}^{2}x_1-\omega_2^{2} x_2,
\end{eqnarray} 
\noindent where $x_1$ and $x_2$ measure the distance to equilibrium (we assume that $\omega_1$,  $\omega_2$, $\omega_{12}$ and $\omega_{21}$ are positive). Using complex notations, we search a solution of the form  $x_j(t)= a_j  \exp\left( i \omega t\right)$ with $j=1,2$. The amplitudes $a_j$ are solution of an homogeneous system
\begin{equation}
\label{s1}
\left[ \begin{array}{cc}
\left( \omega^{2} -\omega_1^{2}\right) & \omega_{12}^{2} \\
\omega_{21}^{2} & \left( \omega^{2} -\omega_2^{2}\right) \\
\end{array} \right]  \left[ \begin{array}{c } a_1 \\a_2 \end{array} \right]=0.
\end{equation}
\noindent This system has a non-zero solution only if the matrix determinant vanishes and we thus get the equation verified by $\omega$ 
\begin{eqnarray}\label{s2}
\left( \omega^{2} -\omega_1^{2}\right) \left( \omega^{2} -\omega_2^{2}\right) -\omega_{12}^{2} 
\omega_{21}^{2}= 0.
\end{eqnarray} 
\noindent This equation has two roots 
\begin{eqnarray}\label{s3}
\omega_{\pm}^{2} = \frac{\omega_1^{2}+\omega_2^{2}}{2}  \pm\sqrt{\left( \frac{\omega_1^{2}-\omega_2^{2}}{2} \right) ^2+\omega_{12}^{2} \omega_{21}^{2} }.
\end{eqnarray} 
\noindent Equation (\ref{s3}) is symmetric as the two oscillators play symmetric roles. In the following, we consider that $\omega_1$ is fixed and we use it as a frequency unit. Then, eq. (\ref{s3}) becomes
\begin{eqnarray}\label{s3a}
\frac{\omega_{\pm}^{2}}{\omega_1^{2}} = \frac{1}{2}\left[\left(1+\frac{\omega_2^2}{\omega_1^{2}} \right)  \pm\sqrt{\left(1-\frac{\omega_2^2}{\omega_1^{2}} \right)^2  +4\kappa^2}
\right], 
\end{eqnarray}
\noindent where we have introduced a dimensionless coupling parameter $\kappa$ defined by
\begin{eqnarray}\label{s3b}
\kappa \equiv  \frac{\omega_{12} \omega_{21}}{\omega_1^2}.
\end{eqnarray}
\noindent  $\omega_{\pm}/\omega_1$ are plotted as a function of $\omega_2/\omega_1$  in fig. \ref{fig1} for various values of $\kappa $. The two frequencies present an avoided crossing, with a width proportional to $\kappa$. For each root $\omega_{\pm}$, the oscillation amplitudes $a_{\pm,1}$ and $a_{\pm,2}$ are given by
\begin{eqnarray}\label{s4}
\frac{ a_{\pm,1}}{a_{\pm,2} } =\frac{\omega_{12}^{2} }{\omega_1^{2}-\omega_{\pm}^{2}}.
\end{eqnarray} 
\noindent At exact resonance, $\omega_2 =\omega_1$, the mixing of the oscillation amplitudes is given by $ a_{\pm,1}/a_{\pm,2} = \mp\sqrt{\omega_{12}^{2}/\omega_{21}^{2}}$, equal to $\mp 1$ in the case of two identical pendulums coupled by a spring. 

\subsection{Avoided crossings in quantum mechanics}

The avoided crossing of $\omega_{\pm}$ is an analogue of the avoided crossings observed in quantum mechanics \cite{Landau65}. If a Hamiltonian depends on a parameter $\lambda$, the energies of the eigenstates present avoided crossings, when plotted as functions of $\lambda$. This is illustrated by the atomic Zeeman effect in the presence of fine (or hyperfine) structure, $\lambda$ being the magnetic field: the energies of levels with the same $m_J$ (or $m_F$) values (where $m_J$ or $m_F$ is the projection of the total angular momentum on the field axis) present avoided crossings. Crossings can be observed if the Hamiltonian presents a symmetry: the energies, which cross each other, are associated to eigenstates belonging to different symmetry classes. As the Zeeman Hamiltonian has the rotation symmetry around the field axis, the levels with different $m_J$ (or $m_F$) values belong to different symmetry classes and crossings between levels with different $m_J$ (or $m_F$) values are observed.

\subsection{Classical analogue of the Rabi oscillation}

We consider the two-oscillator system with the following initial conditions, $x_1(0)=X$, $x_2(0)=0$ with vanishing velocities $dx_1(0)/dt=dx_2(0)/dt=0$, i.e. at $t=0$,  the energy $E_1$ of oscillator $1$ is maximum while the energy $E_2$ of oscillator $2$ vanishes. The evolution of $x_1(t)$ and $x_2(t)$ is exactly given by:
\begin{eqnarray}\label{sr1}
\frac{x_1(t)}{X} &=& \frac{\left(\omega_{+}^2 -\omega_{1}^2 \right) \cos\left(\omega_{-}t \right)-\left(\omega_{-}^2 -\omega_{1}^2 \right) \cos\left(\omega_{+}t \right)}{\omega_{+}^2 -\omega_{-}^2} \nonumber \\
\frac{x_2(t)}{X} &=& \frac{\left(\omega_{+}^2 -\omega_{1}^2 \right)\left(\omega_{-}^2 -\omega_{1}^2 \right)}{\omega_{12}^{2}\left( \omega_{+}^2 -\omega_{-}^2\right) }\left[ \cos\left(\omega_{+}t \right)- \cos\left(\omega_{-}t \right)   \right]. \nonumber \\
\end{eqnarray} 
\noindent  In order to simplify the algebra and to exhibit more clearly the analogy with the Rabi oscillation, we assume that $\omega_{21}=\omega_{12}$ and we introduce the frequency detuning $\delta = \omega_{2} -\omega_{1}$, the mean of the uncoupled oscillator frequencies $\omega_m =\left( \omega_{1}+\omega_{2}\right) /2$ and the equivalent of the Rabi frequency $\Omega_1 = \omega_{12}^2/\omega_m $. If $\delta $ and $\Omega_1 $ are both small with respect to $\omega_m$,  $\omega_{\pm}$ are approximately given by  
\begin{eqnarray}\label{sr2}
\omega_{\pm} &\approx& \omega_m  \pm \frac{1}{2} \sqrt{\delta^2+ \Omega_1^2}
\end{eqnarray} 
\noindent and we can rewrite $x_1(t)$ and $x_2(t)$ 
\begin{eqnarray}\label{sr3}
\frac{x_1(t)}{X} &\approx & \cos\left(\omega_{m}t\right) \cos\left(\sqrt{\delta^2+ \Omega_1^2}\frac{t }{2}\right) \nonumber \\&&-\frac{\delta}{\sqrt{\delta^2+ \Omega_1^2}} \sin\left(\omega_{m}t\right) \sin\left(\sqrt{\delta^2+ \Omega_1^2}\frac{t }{2}\right) \nonumber \\
\frac{x_2(t)}{X} &\approx & \frac{\Omega_1}{\sqrt{\delta^2+ \Omega_1^2}} \sin\left(\omega_{m}t\right) \sin\left(\sqrt{\delta^2+ \Omega_1^2}\frac{t }{2}\right). \nonumber \\
\end{eqnarray} 
\noindent $x_1(t)$ and $x_2(t)$ are both oscillating at the large frequency  $\omega_{m}$ and their oscillation amplitude is slowly modulated at the frequency $\sqrt{\delta^2+ \Omega_1^2}$. If we consider for instance the energy $E_2$ of oscillator $2$ averaged over one period of the fast oscillation, it is given by  
\begin{eqnarray}\label{sr4}
\frac{E_2}{E_1+E_2} =\frac{\Omega_1^2}{\delta^2+ \Omega_1^2}  \sin^2\left(\sqrt{\delta^2+ \Omega_1^2}\frac{t }{2}\right). 
\end{eqnarray} 
\noindent If we consider the Rabi oscillation \cite{RabiPR37,Cohentannoudji88,Loudon83}, with all the population at $t=0$ in level $1$, the population transferred at $t$ in level $2$ is exactly given by the right-hand side of eq. (\ref{sr4}).

\section{Coupled mechanical oscillators with damping}
\label{se3}

\subsection{Coupled equations with damping}

We now add damping terms in the equations of motion of the coupled oscillators and we use anelastic damping terms (see Appendix B), because this choice simplifies the calculations. We assume that the imaginary parts of the coupling terms $\omega_{12}^{2}$ and  $ \omega_{21}^{2} $ are negligible. Then, we have simply to replace $\omega_j^{2} $  by  $\omega_j^{'2} = \omega_j^{2} +i \omega_j\gamma_j$ in eqs. (\ref{s0}) and we assume that the damping is weak $\gamma_j\ll  \omega_j$ ($j=1,2$). Eq. (\ref{s2}) becomes 
\begin{eqnarray}\label{s7}
\left( \omega^{2} -\omega_1^{'2}\right) \left( \omega^{2} -\omega_2^{'2}\right) -\omega_{12}^{2} \omega_{21}^{2}=0.
\end{eqnarray} 
\noindent Thanks to the choice of anelastic damping, this equation is a second degree equation in $\omega^{2}$ whereas this equation would be a fourth degree equation in $\omega$, if we had used viscous damping terms. The solutions of eq. (\ref{s7}) are given by
\begin{eqnarray}\label{s8}
\frac{\omega_{\pm}^{2}}{\omega_1^{2}} = \frac{\omega_1^{'2}+\omega_2^{'2}}{2\omega_1^{2}}   \pm\sqrt{\left(\frac{\omega_1^{'2}-\omega_2^{'2}}{2\omega_1^{2}} \right)^2 +\kappa^2}.
\end{eqnarray} 
\noindent  This result is similar to eqs. (\ref{s3},\ref{s3a}) but, as it involves the square root of a complex number, it less easy to visualize the variations of $\omega_{\pm}$. If the coupling term $\kappa$ is small, the interesting case is close to resonance and we will consider only this case from now on. We make an analytic study in two limiting cases and, afterwards, a numerical one in the general case.

\subsection{Analytic study of the resonance region $\omega_2 /\omega_1 \approx 1$}

The difference of the real parts of the frequencies $\omega_j^{'} $ is small and, in the square root of eq. (\ref{s8}), there are two competing dimensionless terms: the coupling term $\kappa^2 $, which is positive, and the term $\left[ \left(\omega_1^{'2}-\omega_2^{'2}\right)/\left( 2\omega_1^{2}\right)\right] ^2 $ which, exactly at resonance, is negative and equal to $-\left[ \left( \gamma_1-\gamma_2 \right)/\left( 2\omega_1\right)\right]^2$. The behavior is different, depending which term is dominant and it is natural to introduce a critical value $\kappa_{cr}$ of the coupling term defined by 
\begin{eqnarray}\label{s8a}
\kappa_{cr} \equiv \frac{\left|\gamma_1-\gamma_2\right|}{2\omega_1}  =\left|\frac{1}{2Q_1} -\frac{1}{2Q_2} \right|, 
\end{eqnarray} 
\noindent where we have introduced the quality factors of the uncoupled oscillators $Q_1=\omega_1/\gamma_1$ and $Q_2=\omega_1/\gamma_1$ (we may use $\omega_1$ for both oscillators as we study the resonance region $\omega_2 \approx\omega_1$). There are no simple analytic results when $\kappa \approx \kappa_{cr}$ and, in order to get approximate analytic results, we consider two limiting cases $\kappa \gg  \mbox{  or  } \ll\kappa_{cr}$.

\subsubsection{The coupling term is dominant: $\kappa \gg \kappa_{cr}$} 

The mixing of the two oscillators occurs almost as in the absence of damping. The real parts of frequencies $\omega_{\pm}$ present an avoided crossing and, exactly at resonance, they are given by
\begin{eqnarray}\label{s9}
\frac{\omega_{\pm}^{2}}{\omega_1^{2}} \approx \frac{\omega_1^{'2}+\omega_2^{'2}}{2\omega_1^{2}}  \pm\kappa.
\end{eqnarray} 
\noindent This approximate form is valid as long as $\left( \omega_1^{'2}-\omega_2^{'2} \right)^2/\omega_1^{2}\ll 4\kappa^2$: this is verified near the resonance center but not in the wings where the difference $(\omega_1-\omega_2)$ becomes large. When eq. (\ref{s9}) is valid, the two resonance frequencies have the same imaginary part $\approx \left( \gamma_1+\gamma_2\right) /2$. This result can be understood by reference to the case without damping: in the resonance center, the two oscillators are completely mixed.   

\subsubsection{The damping is dominant $\kappa \ll \kappa_{cr}$} 

The behavior is completely different: the real parts of $\omega_{\pm}$ do not present an avoided crossing when the ratio $\omega_2/\omega_1$ varies: $\omega_{+}\rightarrow \omega_1$ (and $\omega_{-} \rightarrow \omega_2$) when the ratio $\omega_2/\omega_1 $ is sufficiently smaller or larger than $1$. We expand the square root appearing in eq. (\ref{s8}) at first order in $\kappa^2$:
\begin{eqnarray}\label{s10}
\frac{\omega_{\pm}^{2}}{\omega_1^{2}} \approx \frac{\omega_1^{'2}+\omega_2^{'2}}{2\omega_1^{2}}  \pm \left[\frac{ \omega_1^{'2}-\omega_2^{'2}}{2\omega_1^{2}} + \kappa^2 \frac{\omega_1^{2}}{\omega_1^{'2}-\omega_2^{'2}} \right], 
\end{eqnarray} 
\noindent from which we deduce $\omega_{\pm}$ 
\begin{eqnarray}\label{s11}
\omega_{+} &\approx & \omega_1 + i \frac{\gamma_1}{2} + \frac{\kappa^2}{4} \times\frac{\omega_1^2}{\omega_1-\omega_2  +i \left( \gamma_1-\gamma_2 \right) /2}\nonumber \\
\omega_{-} &\approx & \omega_2+ i \frac{\gamma_2}{2} - \frac{\kappa^2}{4} \times\frac{\omega_1^2}{\omega_1-\omega_2  +i \left( \gamma_1-\gamma_2 \right) /2}.
\end{eqnarray}
\noindent We have simplified the results by replacing, for instance, $\left( \omega_1^{2}-\omega_2^{2}\right)$ by $2\omega_1 \left(\omega_1-\omega_2\right)$, which is a good approximation because $\omega_2 \approx\omega_1$, and by omitting terms in $\kappa^2\gamma_1$ or $\kappa^2\gamma_2$ which are negligible in the weak damping limit. When the frequency ratio $\omega_2/\omega_1$ varies around $1$, the real and imaginary parts of $\omega_{\pm}$ present resonant variations with Lorentzian line shapes:  a dispersion (respectively absorption) lineshape for the real (respectively imaginary) parts of $\omega_{\pm}$. The full width at half maximum of these resonant variations is the same, equal to $\left( \gamma_1-\gamma_2 \right)$. The damping rate is given by the imaginary parts of $\omega_{\pm}$ and eqs. (\ref{s11}) prove that some damping is transferred from the more damped oscillator to the less damped one. 

If we consider the case of exact resonance, $\omega_2 = \omega_1$, eq. (\ref{s10}) can be simplified and the quality factors $Q_{\pm}$ of the coupled oscillators are given by
\begin{eqnarray}\label{s11a}
\frac{1}{Q_{\pm}}&\equiv & \frac{2{\mathcal{I}}m\left( \omega_{\pm}\right)}{\omega_1} 
\approx \frac{1}{2}\left(\frac{1}{Q_1}  + \frac{1}{Q_2}\right) \nonumber \\
&& \pm \frac{1}{2}\left(\frac{1}{Q_1} -\frac{1}{Q_2}\right) \sqrt{1 -\frac{\kappa^2}{\kappa_{cr}^2} }, 
\end{eqnarray}
\noindent at first order in $1/Q_1 $ and $1/Q_2$. 

From now on, we simplify the discussion by assuming  that $\gamma_2=0$ (i.e. $Q_2 \longrightarrow \infty$): only oscillator $1$ is damped and the damping of oscillator $2$ is solely due to its coupling to oscillator $1$. Then eq. (\ref{s11a}) can be simplified and the quality factor $Q_{-}$ is given by
\begin{eqnarray}\label{s11b}
Q_{-} \approx \frac{1}{\kappa^2 Q_1}.  
\end{eqnarray}
\noindent At first sight, this result is very surprising: the damping of oscillator $2$ induced by coupling to oscillator $1$ decreases when the damping of oscillator $1$ increases. We may nevertheless understand why it is so. The coupling term induces a mixing of the two oscillators which, in the absence of damping, is maximum at exact degeneracy, $\omega_2 = \omega_1$. In the presence of damping of oscillator $1$, the resonance frequency  $\omega_1^{'}$  is complex and the difference  $\left(\omega_1^{'}-\omega_2^{'}\right)$ never vanishes; its minimum modulus is $\gamma_1/2$ and, as the coupling is weak, this distance to degeneracy is sufficient to prevent a strong mixing of the two oscillators. Moreover, when $\gamma_1$ increases, the mixing decreases and, as a consequence, the damping induced on oscillator $2$.  

The resonant variations of the real parts of $\omega_{\pm}$ are also very interesting. In particular, the frequency displacement of the resonance of oscillator $2$ is equal to the real part of $\Delta\omega_{-}\equiv \left( \omega_{-} -  \omega_{2}\right)$. It is larger than its damping rate when $\left|\omega_1-\omega_2\right| > \gamma_1 /2$ and it should be easy to detect this displacement. 

\subsection{Numerical study of the resonance region $\omega_2 /\omega_1 \approx 1$}

We now complement these analytic results by a numerical study. The frequencies and damping rates are referred to $\omega_1$ taken as the frequency unit. We assume that, in the absence of coupling, oscillator $2$ is not damped i.e. $\gamma_2=0$: this choice reduces the number of parameters and enhances the visibility of the damping induced by coupling to oscillator $1$. 

\begin{figure}[h]
\begin{center}
\includegraphics[width= 8 cm]{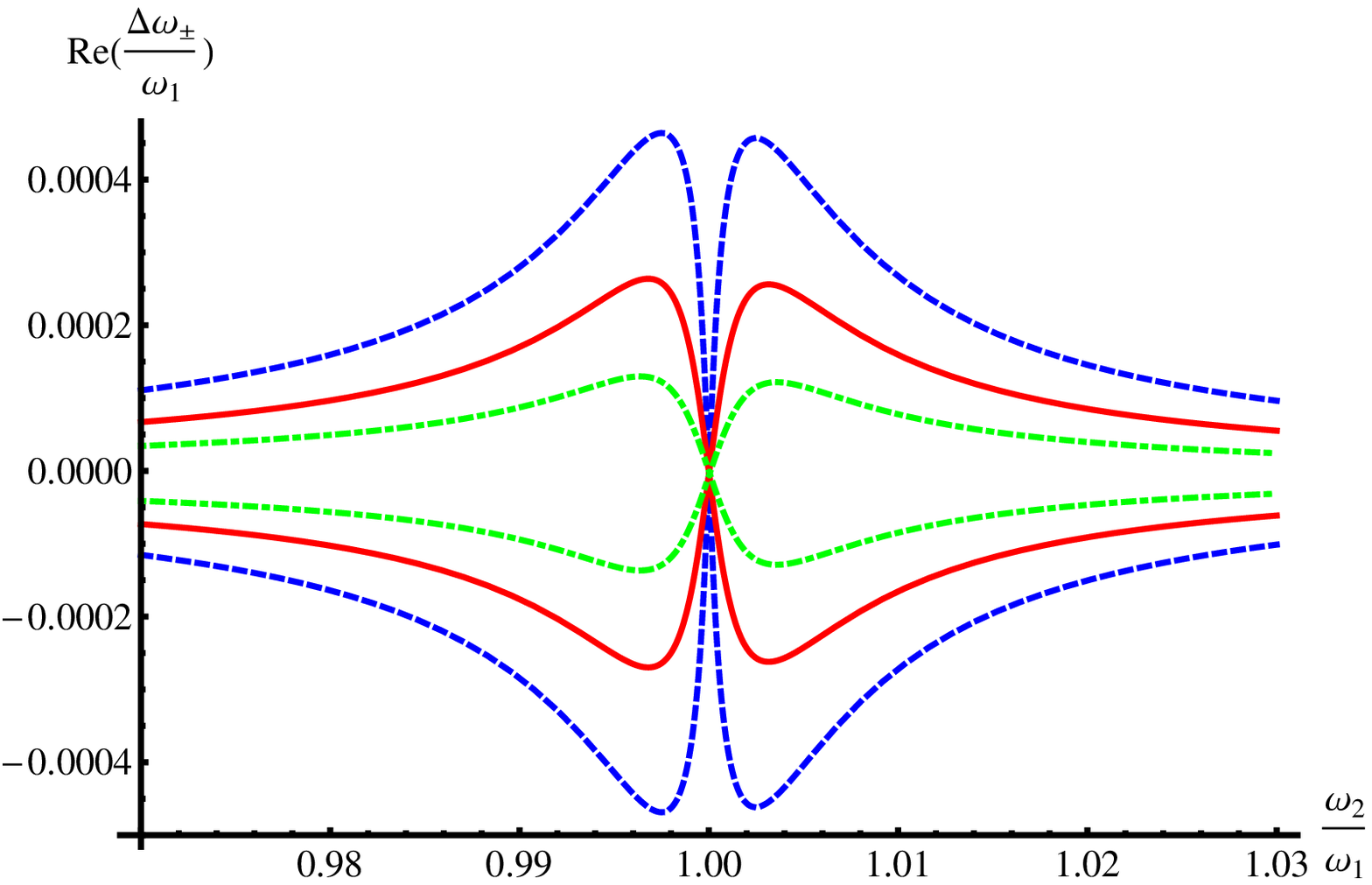}
\includegraphics[width= 8 cm]{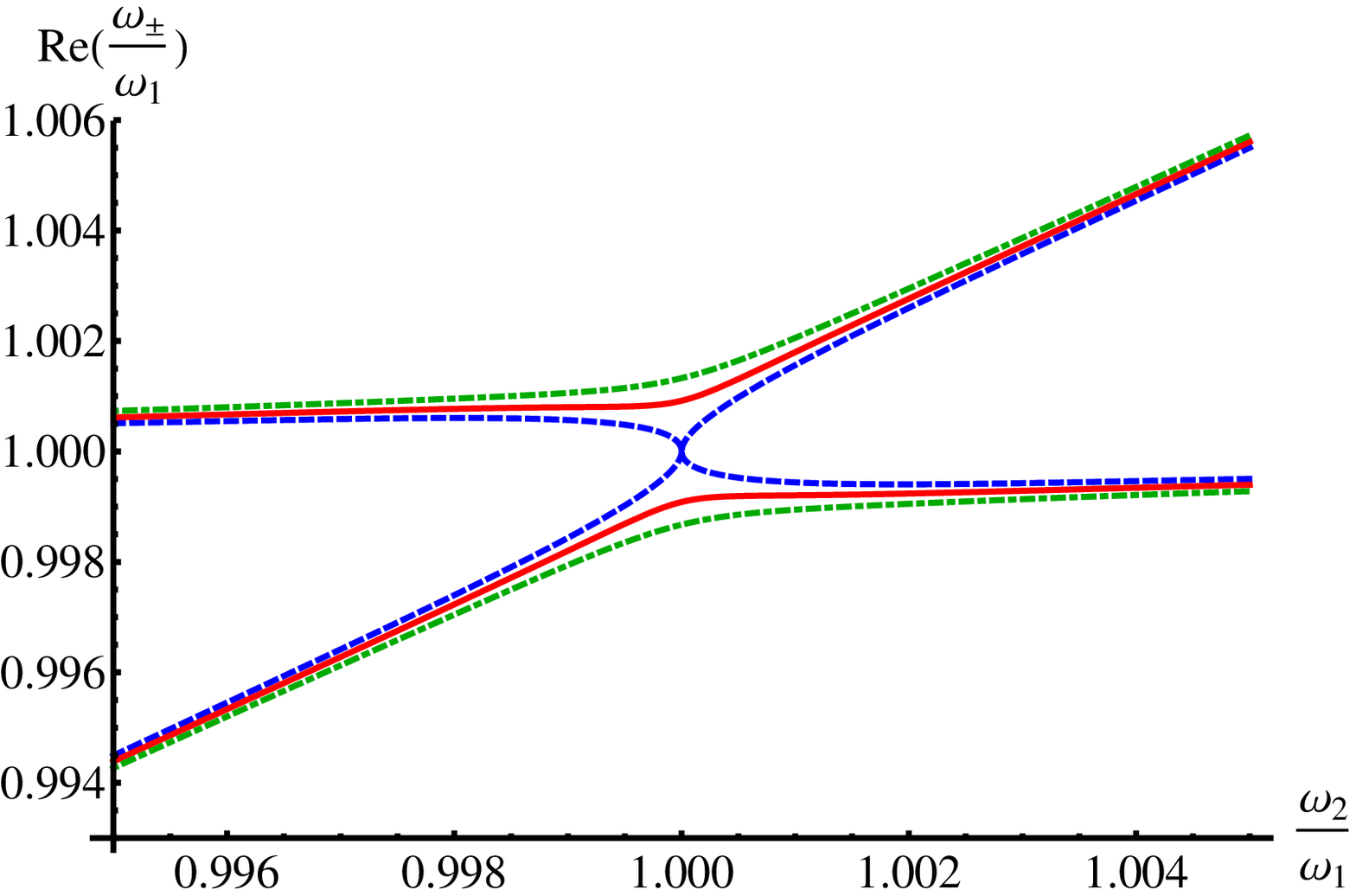}
\caption{(Color online) Effect of the coupling on the real parts of $\omega_{\pm}$. The calculation is done with $\gamma_1/\omega_1= 0.01$ and $\gamma_2=0$ corresponding to the critical $\kappa$-value $\kappa_{cr} = 5\times 10^{-3}$. \\
Upper panel: Plot of the real parts of $\Delta\omega_{+}/ \omega_1= \left( \omega_{+} -  \omega_{1}\right) /\omega_1$  and of $\Delta\omega_{-}/ \omega_1=\left( \omega_{-} -  \omega_{2}\right) /\omega_1$ as a function of $\omega_2/\omega_1$ for the following values of  $\kappa/ \kappa_{cr}$: $0.5$ dot-dashed (green) curves, $0.7$ full (red) curves, and $0.9$ dashed (blue) curves. The curves are close to Lorentz dispersion curves,  with $\Delta\omega_{-}>0$ (respectively $<0$) when $\omega_2 /\omega_1 >1$ (respectively $<1$) and the opposite behavior for $\Delta\omega_{+}$.\\
Lower panel: Plot of the real parts of $\omega_{\pm}$ as a function of $\omega_2/\omega_1$ for the following values of  $\kappa/ \kappa_{cr}$: $1.0$ dashed (blue) curves, $1.1$ full (red) curves,  $1.2$ dot-dashed (green) curves.}\label{fig2} 
\end{center}
\end{figure}

If  $\kappa \leqslant \kappa_{cr}$, the values of $\omega_{+}$ and $\omega_{-}$ are approximately given by eqs. (\ref{s11}) and  there is no avoided crossing. We have plotted the real parts of $\Delta\omega_{\pm}/ \omega_1\equiv \left( \omega_{\pm} - \omega_1\right) /\omega_1$ as a function of $\omega_2 /\omega_1$ in the upper panel of  fig. \ref{fig2}: as predicted by our analytic results, these curves are close to Lorentz dispersion curves. 

If  $\kappa > \kappa_{cr}$, there is an avoided crossing at resonance and we have plotted  the real parts of $ \omega_{\pm}/\omega_1$  as a function of $\omega_2 /\omega_1$ in the lower panel of fig. \ref{fig2}: the shape of these curves is complicated when $\kappa$ is only slightly larger than $\kappa_{cr}$ but, when $\kappa$ increases, these curves rapidly become very similar to those calculated in the absence of damping and presented in fig. \ref{fig1}.

\begin{figure}[h]
\begin{center}
\includegraphics[width= 8 cm]{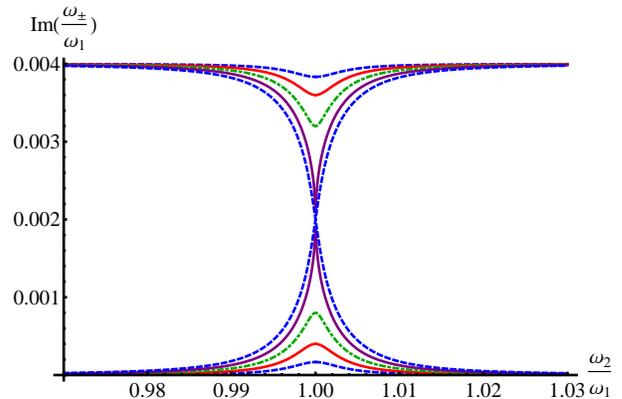}
\caption{ Plot of the imaginary parts of $\omega_{\pm}/\omega_1$ as a function of the ratio $\omega_2/\omega_1$.  $\gamma_1/\omega_1= 0.01$ and $\gamma_2=0$, corresponding to $\kappa_{cr} = 5\times 10^{-3}$. The different curves correspond to $\kappa/ \kappa_{cr} = 0.4$ dashed (blue) curves; $0.6$ full (red) curves; $0.8$ dot-dashed (green) curves; $1$ full (violet) curves and  $1.2$ dashed (blue) curves. }
\label{fig3}
\end{center}
\end{figure} 

Figure \ref{fig3} presents the variations of   ${\mathcal{I}}m\left( \omega_{\pm}\right)/\omega_1$ as a function of $\omega_2 /\omega_1$ for a series of $\kappa$-values. These plots are close to Lorentz absorption curves when $\kappa$ is below $\kappa_{cr}$  while, when $\kappa \geqslant \kappa_{cr}$, the imaginary parts cross each other when $\omega_2 = \omega_1$. 

\subsection{Comparison with quantum mechanics}

The two regimes discussed above appear to be fully similar to those observed in quantum mechanics when a discrete state is coupled to a continuum. As recalled in the introduction, there are two well-known limiting cases of this dynamics: the Weisskopf-Wigner exponential decay of the discrete state \cite{WW1930a,WW1930b} is a good approximation when the discrete state is weakly coupled to the continuum while the Rabi oscillation between two discrete states \cite{RabiPR37} is a good approximation if the continuum width is negligible. The continuous transition between these two regimes, first discussed by C. Cohen-Tannoudji and P. Avan \cite{Cohentannoudji77} in 1977, is described in detail in the book ``Atom-photon interactions'', by C. Cohen-Tannoudji, J. Dupont-Roc and G. Grynberg \cite{Cohentannoudji88}. We cannot reproduce here this discussion but we may summarize its results. The important quantities are the width $w_0$ of the continuum and the Rabi frequency $\Omega_1$: 

a) if $w_0 \gg \Omega_1$, the dynamics is well described by the Weisskopf-Wigner exponential decay with the decay rate approximately given by the Fermi golden rule. The density of states appearing in the Fermi golden rule is inversely proportional to the width $w_0$ and, as a consequence, the decay rate of the discrete state is also $\propto 1/w_0$. As $w_0$ gives the decay rate of a continuum wavepacket, the decay rate of the discrete state decreases when the decay rate of a continuum wavepacket increases. In addition, the coupling to the continuum also induces an energy shift, which is usually difficult to measure and this difficulty explains why it is rarely discussed. However, this shift is famous in the case of the Lamb shift, first discovered in the $n=2$ level of of hydrogen \cite{LambPR47} and its discovery has played a very important role in the development of Quantum Electro-Dynamics. This shift exists also in molecular predissociation (see ref. \cite{ChildCJP75} and references therein). 

b) in the opposite case, if $w_0 \ll \Omega_1$, the dynamics is well described by a Rabi oscillation between the discrete state and the narrow continuum which behaves as a discrete state, at least for timescales smaller than $h/w_0$ (where $h$ is the Planck constant). 

Finally, by varying the relative magnitude of $\Omega_1$ and $w_0$, one can observe a continuous transition between these two limiting cases  \cite{Cohentannoudji88}. Here is an application of these ideas:  we consider an atom in its ground state coupled by a resonant laser to one of its excited states. Because of spontaneous emission, the excited state has a finite lifetime $\tau$ and it is a continuum of finite width $w_0\sim h/\tau$. The discussion is simpler if we may neglect spontaneous emission toward the ground state. If the Rabi frequency $\Omega_1$ is weak, $\Omega_1 \tau \ll 1$,  the effect of the laser is to transfer ground state atoms in the excited state from which they never come back: this means that the laser has given a finite lifetime to the ground state and this lifetime, proportional to $1/(\Omega_1^2 \tau) $, increases when the excited lifetime proportional to $\tau$ decreases. In the opposite case of a strong coupling, $\Omega_1 \tau \ll 1$, the dynamics is described by a Rabi oscillation with a period inversely proportional to $\Omega_1$ and both states have the same lifetime equal to $2\tau $.

\section{Concluding remarks}
\label{se4}

In this paper, we have first recalled the coupling of two mechanical oscillators in the absence of damping. In a second step, we have added the effect of damping and we have shown that the behavior is very different, depending on the relative magnitude of the coupling and damping terms:

\begin{itemize} 

\item  the case when the coupling dominates the damping is classic. Then, the mixing of the two oscillators can be treated almost as in the absence of damping. The resonance frequencies are repelled by the coupling and the two oscillators are mixed by the coupling. As a consequence, the damping is shared, proportionally to the mixing induced by the coupling. 

\item the original case occurs when the difference of damping rates dominates the coupling. Then, the coupling has weaker effects on the frequencies and on the damping rates of the two oscillators. If the frequency of one oscillator is swept close to resonance, the real and imaginary parts of the frequencies of the coupled oscillators present resonant variations with Lorentzian lineshapes, a dispersion lineshape for the real parts of the frequencies and an absorption lineshape for their imaginary parts. Some damping is transferred from the more damped oscillator to the less damped one with a surprising result: the damping transferred decreases when the damping of the more damped oscillator increases. 
 
\end{itemize}

These two regimes are very similar to what occurs in quantum mechanics with the continuous transition between the Rabi oscillation regime and the Weisskopf-Wigner exponential decay when a discrete state is coupled to a continuum. 

\section{Appendix A: example of coupled mechanical oscillators}
\label{se5}

\subsection{Two pendulums coupled by a spring}
\label{se51}

\begin{figure}[h!]
\includegraphics[width=8.0 cm]{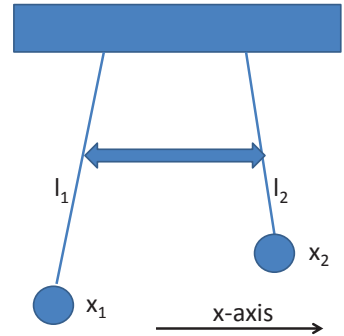}
\caption{ Schematic drawing of two simple pendulums coupled by a spring represented by the large horizontal arrow. }
\label{fig4}
\end{figure}

We first consider the case of two simple pendulums of masses $m_1$ and $m_2$ and of lengths $l_1$ and $l_2$ (see fig. \ref{fig4}). We note $x_1$ and $x_2$ their displacements from equilibrium and we assume a coupling force proportional to the difference $(x_1-x_2)$. The equations of motion are
\begin{eqnarray}\label{r1}
m_1 \frac{d^{2}x_1}{dt^{2}} &=& -k_1 x_1 -k_{12}(x_1-x_2),\nonumber \\
m_2 \frac{d^{2}x_2}{dt^{2}} &=& -k_2 x_2 -k_{21}(x_2-x_1),
\end{eqnarray} 
\noindent where $k_1= m_1g/l_1$ and $k_2=m_2g/l_2$, $g$ being the acceleration of gravity. 
Because of the equality of action and reaction,  $k_{12}= k_{21}$. Noting $\omega_1^{2} = (k_1+ k_{12})/m_1$, $\omega_2^{2}  = (k_2+ k_{12})/m_2$, $\omega_{12}^{2} = k_{12}/m_1$ and $\omega_{21}^{2} = k_{12}/m_2$, we get
\begin{eqnarray}\label{r3}
\frac{d^{2}x_1}{dt^{2}} &=& -\omega_1^{2} x_1 +\omega_{12}^{2} x_2, \nonumber \\
\frac{d^{2}x_2}{dt^{2}} &=&  +\omega_{21}^{2}x_1-\omega_2^{2} x_2.
\end{eqnarray} 
\subsection{Double pendulum}
\begin{figure}[h]
\begin{center}
\includegraphics[width=8.0 cm]{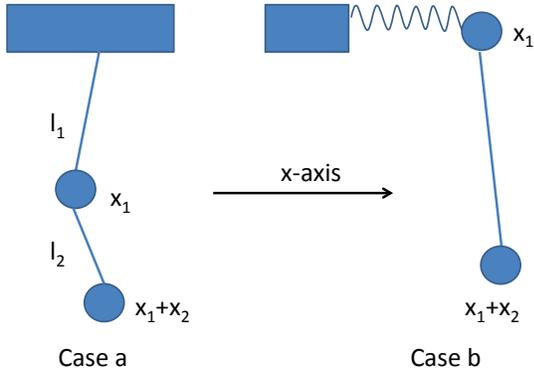}
\caption{ Schematic drawing of two different double pendulums with the top pendulum being either a simple pendulum (case a) or an elastic pendulum (case b).}
\label{fig5}
\end{center}
\end{figure}
Two possible arrangements of a double pendulum are shown in fig. \ref{fig5}. The equations of motion are

\begin{eqnarray}\label{r4}
m_1 \frac{d^{2}x_1}{dt^{2}} &=& -k_1 x_1 +k_{2}x_2,\nonumber \\
m_2 \frac{d^{2}\left( x_1+x_2\right) }{dt^{2}} &=& -k_2 x_2. 
\end{eqnarray} 

\noindent In case a of fig. \ref{fig5}, $k_1= m_1g/l_1$ while, in case b,  $k_1$ is the elastic constant of the spring and, in both cases, $k_2=m_2g/l_2$. Thanks to the first equation, we eliminate $d^{2}x_1/dt^{2}$ from the second one and we get

\begin{eqnarray}\label{r5}
\frac{d^{2}x_1}{dt^{2}} &=& -\frac{k_1}{m_1} x_1 +\frac{k_2}{m_1}x_2,\nonumber \\
\frac{d^{2}x_2}{dt^{2}} &=& \frac{k_1}{m_1}x_1- \frac{k_2\left( m_1+m_2\right) }{m_1m_2} x_2. 
\end{eqnarray} 

\noindent Noting $\omega_1^{2} = k_1/m_1$, $\omega_2^{2} = k_2(m_1+m_2)/(m_1m_2)$, $\omega_{12}^{2} =k_{2}/m_1$ and $\omega_{21}^{2} = k_{1}/m_1=\omega_1^{2}$, we also get eqs. (\ref{r3}). In this case, the coupling parameter $\kappa$ defined by eq. (\ref{s3b}) is equal to $\kappa =\sqrt{k_2/k_1}$ and, if we assume resonance $\omega_2=\omega_1$, $\kappa =\sqrt{m_2/(m_1+m_2)}$.

\section{Appendix B: damping of mechanical oscillators}
\label{se6}

The damping of a mechanical oscillator is treated in most textbooks. The equation of motion is linear in two main cases: i) a damping force proportional to the velocity due, for instance, to the friction on a fluid in the Stokes regime \cite{Stokes1850}; ii) an anelastic behavior of the spring \cite{ZenerPR37,ZenerPR38}, i.e. a restoring force which is not in phase with the displacement. Anelastic effect is described by an extension of Hooke's law \cite{SaulsonPRD90}, with a force proportional to $\left[ 1+i\phi(\omega) \right] x$ in complex notations. Here $x$ is the displacement with respect to equilibrium and $\phi(\omega)$ the phase shift between the force and the displacement for an oscillation at a frequency $\omega$. The equations of motion are 
\begin{eqnarray}\label{q1}
\frac{d^{2}x}{dt^{2}} &=& -\omega_0^{2} x -\gamma \frac{dx}{dt} \mbox{ (viscous)},\nonumber \\
\frac{d^{2}x}{dt^{2}} &=& -\omega_0^{2}\left[ 1+i\phi(\omega)\right] x \mbox{  (anelastic)}.
\end{eqnarray} 
\noindent  $x(t) \propto \exp\left( i \omega t\right) $ is solution if $\omega$ verifies the following equations
\begin{eqnarray}\label{q3}
\omega^{2} -i \gamma\omega -\omega_0^{2} &=&0 \mbox{  (viscous)}, \\ 
\omega^{2} -\omega_0^{2} \left[ 1+i\phi(\omega)\right] &=&0 \mbox{  (anelastic)}.
\end{eqnarray} 
\noindent  $\omega$ is then given by 

\begin{eqnarray}\label{q4}
 \omega &=& \pm\sqrt{\omega_0^{2}- \frac{\gamma^{2}}{4}}+i \frac{\gamma}{2}   
 \approx\pm\omega_0 + i \frac{\gamma}{2} \mbox{  (viscous)}, \label{q4a} \\
 \omega & = &  \pm\sqrt{\omega_0^{2}\left[ 1+i\phi(\omega_0)\right] }
 \approx  \pm\omega_0\left[ 1+i\frac{\phi(\omega_0)}{2}\right] \label{q4b} \mbox{  (anelastic)}, \nonumber \\
\end{eqnarray}

\noindent with the approximate forms valid in the weak-damping limit, $\gamma \ll  \omega_0$ or $\left|\phi(\omega_0)\right| \ll 1$. Moreover, to insure damping and not amplification, $\phi(\omega)$ must be an odd function of $\omega$ and we have chosen $\omega$ and $\phi(\omega)$ both positive. Both damping mechanisms lead to an exponential decrease of the oscillation amplitude with  $ x(t) \approx x(0) \exp\left(-t/\tau\right) \cos\left(\omega_0 t\right)$. The decay time constant is equal to $\tau =2/\gamma$ or $\tau = 2/\left( \omega_0\phi(\omega_0)\right)$. The resonance quality factor defined by $Q \equiv  \omega\tau /2$ is equal to $Q = {\mathcal{R}}e(\omega)/(2{\mathcal{I}}m(\omega)) = \omega_0/\gamma$ or $Q = 1/\phi(\omega_0)$. If we add on the right hand side of eqs. (\ref{q1}) a driving term $b\exp\left( i \omega t\right)$, the steady state regime is given by

\begin{eqnarray}\label{q5}
x(t) &=& \frac{b\exp\left( i \omega t\right)}{\omega^{2} -i \gamma\omega -\omega_0^{2}}   \mbox{  (viscous)}, \\ 
x(t) &=& \frac{b\exp\left( i \omega t\right)}{\omega^{2} -\omega_0^{2} \left( 1+i\phi(\omega_0)\right)} \mbox{  (anelastic)}.
\end{eqnarray} 

\noindent The two resonances have very similar lineshapes, quasi-Lorentzian in the viscous case and exactly Lorentzian in the anelastic case, with a maximum of the amplitude for  $\omega \approx \omega_0$ and a resonance full width equal to $\gamma$ or $\omega_0\phi(\omega_0)$. The difference, which appear in the far wings, are of minor importance as the interest is focused on the resonance core. As a conclusion, both mechanisms are almost equivalent if $\gamma=\omega_0\phi(\omega_0)$. We use anelastic damping in the calculations of section \ref{se3}, because it simplifies the algebra, and we replace $\omega_0\phi(\omega_0)$ by $\gamma$ so that the equations look closer to the viscous damping discussed in most textbooks.


\newpage

\end{document}